# Affective Ludology, Flow and Immersion in a First-Person Shooter: Measurement of Player Experience

## Lennart Nacke and Craig A. Lindley
Blekinge Institute of Technology, School of Computing
**lennart.nacke@acm.org**     **craig.lindley@bth.se**

## Abstract

Gameplay research about experiential phenomena is a challenging undertaking, given the variety of experiences that gamers encounter when playing and which currently do not have a formal taxonomy, such as flow, immersion, boredom, and fun. These informal terms require a scientific explanation. Ludologists also acknowledge the need to understand cognition, emotion, and goal-oriented behavior of players from a psychological perspective by establishing rigorous methodologies. This paper builds upon and extends prior work in an area for which we would like to coin the term "affective ludology." The area is concerned with the affective measurement of player-game interaction. The experimental study reported here investigated different traits of gameplay experience using subjective (i.e., questionnaires) and objective (i.e., psychophysiological) measures. Participants played three *Half-Life 2* game level design modifications while measures such as electromyography (EMG), electrodermal activity (EDA) were taken and questionnaire responses were collected. A level designed for combat-oriented flow experience demonstrated significant high-arousal positive affect emotions. This method shows that emotional patterns emerge from different level designs, which has great potential for providing real-time emotional profiles of gameplay that may be generated together with self-reported subjective player experience descriptions.

## Introduction

The research field of game science with a focus on experimental research is growing. Prior studies of digital games have often focused on the negative effects digital gaming, such as violent content and its impact (Bushman & Anderson, 2002) or addiction to playing (Grüsser, Thalemann, & Griffiths, 2007). However, there has been a recent shift of focus toward trying to understand aspects central to gameplay experience (Komulainen, Takatalo, Lehtonen, & Nyman, 2008; Mäyrä & Ermi, 2005; Nacke, Drachen, et al., 2009). The primary concern of most research here is to gain a more thorough understanding of loosely defined subjective experiences, such as immersion (Jennett, et al., 2008), presence (Slater, 2002) and flow (Cowley, Charles, Black, & Hickey, 2008). Not only do these terms currently lack well-accepted common meanings, but also clear and testable definitions of these theoretical constructs would be invaluable for game designers, since they are considered to be the holy grail of digital game design. If we had recipes or formulas for creating immersive or flow experiences, these would certainly rather quickly become an industry standard for most games. The shift we can witness in recent years in the games industry is that more focus is put on the players and how to design around an intensely pleasurable player experience. For gaining a better holistic understanding of player experience, we ultimately have to research motivation (Ryan, Rigby, & Przybylski, 2006; Tychsen, Hitchens, & Brolund, 2008), emotions (Mandryk, Atkins, & Inkpen, 2006; Ravaja, Turpeinen, Saari, Puttonen, & Keltikangas-Järvinen, 2008), cognition (Lindley, Nacke, & Sennersten, 2007; Lindley & Sennersten, 2006) and affect of players (Gilleade, Dix, & Allanson, 2005; Hudlicka, 2008; Sykes & Brown, 2003).

*Player Experience and Affective Ludology*

Gameplay or player experience research is evolving – alongside industry advancements concerning behavioral data recording of players (Drachen, Canossa, & Yannakakis, 2009) – to be a fundamental concept in an expanding field of work with a strong empirical research focus. Thus, it applies interdisciplinary research methods from human-computer interaction, computer science, neuroscience, media studies, psychophysiology and psychology to name a few. With this comes a necessary shift in ludology, which has in the past been focused primarily on analyzing games (Juul, 2005; Tychsen, Hitchens, Brolund, & Kavakli, 2006) or establishing a design vocabulary (Church, 1999; Hunicke, LeBlanc, & Zubek, 2004), taxonomies (Lindley, 2003) and ontologies (Zagal, Mateas, Fernandez-Vara, Hochhalter, & Lichti, 2005). Ludology now acknowledges the need to understand cognition, emotion, and goal-oriented behavior of players from a psychological perspective by establishing more rigorous methodologies (Lindley, Nacke, & Sennersten, 2008; Ravaja, et al., 2005).

It has long been argued that a comprehensive theory of game design and development should incorporate multidisciplinary approaches informed by cognitive science, attention and schema theory, emotion and affect, motivation and positive psychology (Lindley & Sennersten, 2008). The improvement of scientific methodologies for studying players and games will not only help us understand the aesthetics of digital games better, but also the underlying processes involved in creating individual player experiences (Tychsen & Canossa, 2008). Building on a foundation laid out by the seminal works of Klimmt (2003), Ravaja (2004), Mandryk and Inkpen (Mandryk & Inkpen), Hazlett (2006), and Mathiak and Weber (2006) the term *affective ludology* is proposed by Nacke (2009) for referring to the field of research which investigates the affective interaction of players and games (with the goal of understanding emotional and cognitive experiences created by this interaction). In this context, players can be seen as biological systems that develop out of the interaction of several complex variables, constituting the human processes known as emotion and cognition.

The psychophysiological study reported in this paper takes a step forward into affective ludology with a focus on First-Person Shooter (FPS) games, which aim at providing an immersive gameplay experience for players by removing self-representations (such as avatars) and putting players in first-person perspective. In a FPS game, players can fully identify with game characters represented only by weapons or hands, shown as virtual prostheses that reach into the game environment (Grimshaw, 2008). This means that players virtually turn into game characters in a FPS game, since they feel like they are acting directly in the virtual game world. In addition to the FPS perspective, the consequence and meaning of player action within the environment and its impact on gameplay greatly influence the feeling of immersion (Ermi &

Mäyrä, 2005; Jennett, et al., 2008; McMahan, 2003). The study of FPS games may simplify the investigation of immersion, flow and presence by removing issues of potential identification with a character viewed from a third-person perspective. While gameplay experience may consist of many different factors, the most discussed ones in related literature are immersion, flow, and presence. In this study, we aim at getting a better understanding of these factors together with an interpretation of the physiological responses of players when playing a game.

### Immersion

A qualitative study conducted by Brown and Cairns (2004) analyzed players' feelings toward their favorite game and led them to propose three gradual and successive levels of player immersion: (1) engagement, (2) engrossment, and (3) total immersion. The latter level (3) is used interchangeably with the concept of presence; a state facilitated by feelings of empathy and atmosphere, which links immersion to factors of graphics, plot, and sounds in addition to emergent gameplay (since visual, auditory or mental elements are mentioned in this context). In a similar way, Jennett et al. (2008) give a conceptual overview of immersion and define it as a gradual, time-based, progressive experience that includes the suppression of all surroundings, together with focused attention and involvement in the sense of being in a virtual world. While it is plausible to see immersion as a gradual phenomenon that increases over playing time, these studies show that lack of a clear definition for presence and immersion can cause the terms to be used interchangeably for phenomena which may not be the same.

Ermi and Mäyrä (2005) subdivided immersion into three distinct forms in their SCI model: (1) sensory, (2) challenge-based and (3) imaginative immersion. "Sensory immersion" concerns the audiovisual execution of games. This dimension of immersion is easily recognizable, since it can be strengthened by intensifying its components, such as creating more compelling high-definition graphics or playing with a much larger video screen or with a surround sound speaker system (Ivory & Kalyanaraman, 2007). "Imaginative immersion" comes close to one part of the immersion definition used by Brown and Cairns (2004), describing absorption in the narrative of a game or identification with a character, which is understood to be synonymous with feelings of empathy and atmosphere. However, atmosphere might be a mix of imaginative immersion and sensory immersion. Hence, the use of this term in the study conducted by Ermi and Mäyrä (2005) raises the need for a clearer definition of the concept of atmosphere. Imaginative immersion is held to be most prominent in role-playing games (Tychsen, Hitchens, Brolund, McIlwain, & Kavakli, 2007). The dimension of challenge-based immersion is very close to what Csíkszentmihályi (1975, 1990) describes as the flow experience. Challenge-based immersion describes the emergent gameplay experience of a player balancing his abilities against the challenges of the game in so far as gameplay is related to motor and mental skills. Challenges in this definition can include different mixtures of physical and mental performance requirements.

Finally, in the study of Ermi and Mäyrä (2005), the game *Half-Life 2* (Valve Corporation, 2004) was ranked highest in all dimensions of the SCI model, thus making it a good candidate for studies investigating immersion. The study reported here, based on *Half-Life 2*, shows a fluid transition between experiential concepts of immersion and flow.

*Flow*

The flow model was introduced by Csíkszentmihályi (1975) based upon his studies of the intrinsically motivated behavior of artists, chess players, musicians and sports players. This group was found to be rewarded by executing actions per se, experiencing high enjoyment and fulfillment in activity in itself (rather than goals of future achievement). Csíkszentmihályi describes flow as the "holistic sensation that people feel when they act with total involvement". Logically, one could see immersion as a precondition for flow, since immersion involves a loss of a sense of context, while flow describes a level of complete involvement. Csíkszentmihályi specified flow as consisting of several characteristics: balance of challenge and skills, clear goals, explicit feedback, indistinct sense of time, loss of self-consciousness, feeling of enjoyment and control in an autotelic (i.e., self-sufficient) activity.

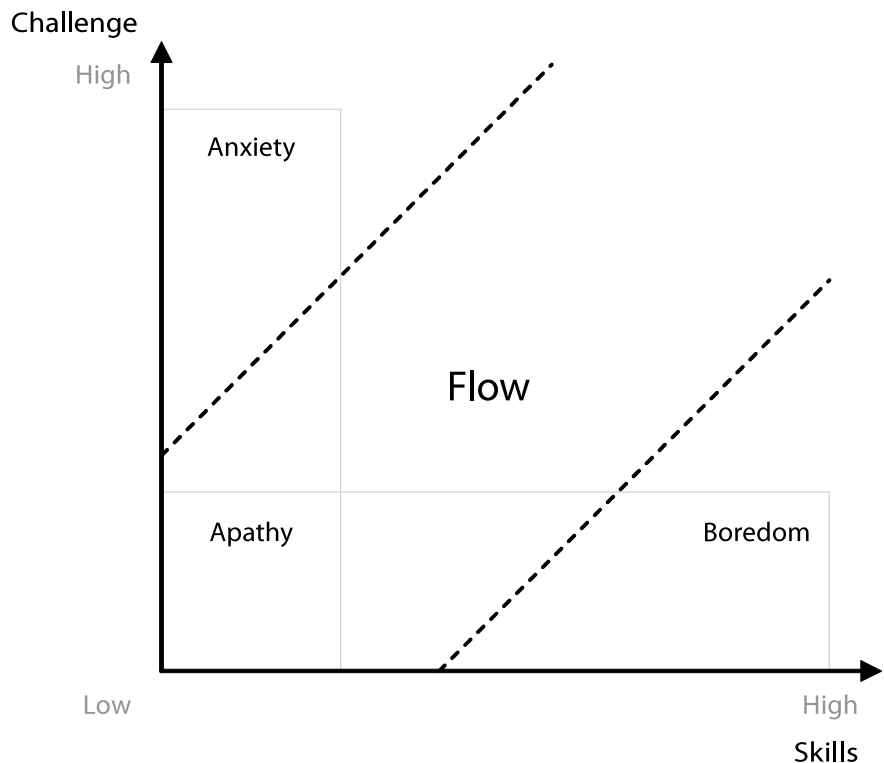

*Figure 1: The two-dimensional four-channel model of flow based on Csíkszentmihályi (1975) and Ellis, Voelkl, and Morris (1994).*

The original flow model was revised by Ellis, et al. (1994) into a four-channel model, shown in Figure 1 which is used most commonly for describing games and gameplay experience. Defining the balance of skills and challenges is often fuzzy, which led Chen (2006) to propose different "flow zones" for hardcore and novice players and an optimal intersection, within which the experience converges towards an optimal match of challenges and abilities.

However, a study by Novak, Hoffman, and Yung (2000) shows that there are many different concepts used for studying flow. They report 16 flow studies between 1977 and 1996, which all use different concepts and definitions of flow. The only commonly used questionnaire, the flow state scale (Jackson & Marsh, 1996), was designed for sports research. It was assessed by Kivikangas (2006) as being usable for game research. In more recent efforts of the EU-funded

Fun of Gaming (FUGA) project, another well-suited scale was developed as part of a Game Experience Questionnaire (GEQ) (IJsselsteijn, Poels, & de Kort, 2008). Kivikangas (2006) was also one of the first to investigate correlations between psychophysiological measures and flow experience, but his results supported no significant relationship between flow state scale and psychophysiological measures.

### Presence

The concept of presence can be discussed briefly in relation to immersion (Slater & Wilbur, 1997), but it is often defined as a state of mind (of being transferred to an often virtual location) rather than a gradual timely experience (Lombard & Ditton, 1997). However, since Zahorik and Jenison (1998) propose to investigate the coupling of perception and action to find an accurate definition of presence, further research is needed to justify a more precise differentiation between presence and immersion. However, spatial presence is a better defined, two-dimensional construct in which the core dimension is the sensation of a physical location in a virtual spatial environment and the second dimension depicts perceived action possibilities (i.e., users only perceive possible actions that are relevant to the virtual mediated space) (Wirth, et al., 2007).

### *Affective and Psychophysiological Measurements*

Norman (2004) makes a clear distinction between emotion and affect, defining emotion as consciously experienced affect, which allows us to identify who or what caused our affective response and why. Affect on the contrary is defined as a discrete, conscious, subjective feeling that contributes to and influences players' emotions. Both emotions and affect are a vital part of player experience, ultimately motivating the cognitive decisions made during gameplay. Psychophysiological research suggests that at least some emotional states can be quantitatively characterized via measurement of physiological responses. Psychophysiology per definition investigates the relationships between psychological manipulations and resulting physiological responses, measured in living organisms (in our case human players) to promote understanding of mental and bodily processes and their relation to each other (Andreassi, 2000). Specific types of measurement of different responses are not *per se* trustworthy signs of well-characterized feelings (Cacioppo, Tassinary, & Berntson, 2007a); a *de rigueur* cross-correlation of all measurements is fundamental to discover the emotional meaning of different patterns in the responses. Furthermore, the often described many-to-one relation between psychological processing and physiological response (Cacioppo, Tassinary, & Berntson, 2007b) allows linking psychophysiological measures to a number of psychological structures (e.g., attention, emotion, information processing). Using a response profile for a set of physiological variables enables scientists to go into more detail with their analysis and allows a better correlation of response profile and psychological event (Cacioppo, et al., 2007b; Mandryk, 2008; Ravaja, 2004). The central concern here is analyzing patterns of measurement characteristics for a set of different measures with subjective characterizations of experience such as emotion and feelings (e.g., the feeling of immersion in gameplay).

Facial electromyography (EMG) is a direct measure of electrical activity involved in facial muscle contractions; EMG provides information on emotional expression via facial muscle activation (even though a facial expression may not be visually observable) and can be considered as a useful external measure for hedonic valence (degree of pleasure/displeasure) (Lang, 1995; Russell, 1980). Positive emotions are indexed by high activity in the *Zygomaticus*

*Major* (ZM, cheek muscle) and *Orbicularis Oculi* (OO, periocular muscle) regions. In contrast to this, negative emotions are associated with high activity in the *Corrugator Supercilii* (CS, brow muscle) regions.

This makes facial EMG suitable for mapping emotions to the valence dimension in the two-dimensional space described in the circumplex affect model (Lang, 1995; Russell, 1980). The *valence* dimension reflects the degree of pleasantness of an affective experience. The other dimension, the *arousal* dimension, depicts the activation level linked to an emotionally affective experience, ranging from calmness to extreme excitement. In this kind of dimensional theory of emotion, emotional categories found in everyday language, such as happiness, joy, depression, and anger, are interpreted as correlating with different ratios of valence and arousal, hence being mappable within a two-dimensional space defined by orthogonal axes representing degrees of valence and arousal, respectively. For example, depression may be represented by low valence and low arousal, while joy may be represented by high valence and high arousal.

Arousal is commonly measured using Electrodermal Activity (EDA), also known as galvanic skin response or skin conductance (Boucsein, 1992; Lykken & Venables, 1971). The conductance of the skin is directly related to the production of sweat in the *eccrine* sweat glands, which is entirely controlled by the human sympathetic nervous system. Increased sweat gland activity is directly related to EDA. Hence, measuring both EDA and EMG provides sufficient data to provide an interpretation of the emotional state of a player.

This paper describes a study investigating correlations between subjectively reported gameplay experience (using questionnaires) and objectively measured player responses (using EMG and EDA) within gameplay, in order to provide cross-validated descriptions of the emotional experience of players during gameplay[1]. The overarching goal is to establish and validate a method that can precisely assess emotional modulations during gameplay (potentially in real-time), for players of FPS games and other genres.[2] The experiment reported in this paper was conducted in February 2008 in the games laboratory of a technical university in Sweden. Although this paper is limited to the description of EMG, EDA and questionnaire data, future analyses will take into account additional data collected (such as eye tracking and electroencephalographic [EEG] data). In the following, we will give an overview of our experimental methodology, and will then continue to report our results. The findings will be discussed and a prognosis for future work will be given.

## Method

The overarching objective of this experimental study was to help us understand how we can use EMG and EDA to measure affective gameplay experience for different game level designs. Thus, we designed three *Half-Life 2* (Valve Corporation, 2004) game modifications (mods) in a highly atmospheric horror setting to guarantee an affective experience. Male students from a technical university played these mods that were specifically designed to test experiential gameplay constructs (and iteratively refined in their design by game testing for half a year). Levels were designed for *immersion*, *boredom*, and *flow*, with each level design modality being played one time. Physiological responses were measured continuously during each play session for each experimental participant (as objective or external measures), while questionnaire data (assessing subjective individual responses) was collected for each participant in each modality after playing a game level.

### Design of the Game Levels and the Experimental Study

The *Half-Life 2* mod game levels were designed addressing the independent variable or within-subject factor of *player experience* in three levels *boredom*, *immersion* and *flow*.

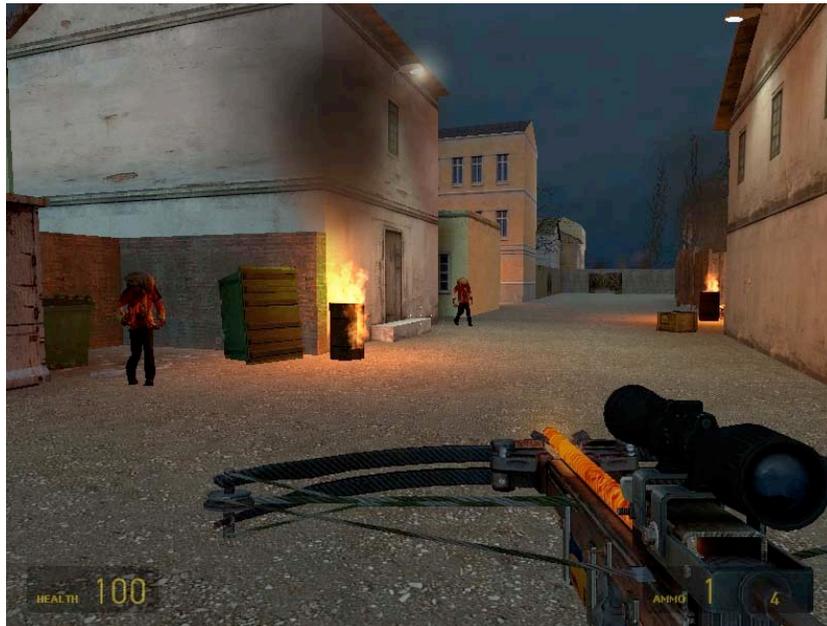

*Figure 2: In-game screenshot for the level designed around the experiential concept of game boredom.*

Dependent variables were EMG, EDA, and subjective GEQ responses. Level design (see Figure 2 was done iteratively, refining the game levels in each design cycle based on feedback from players, game level designers and researchers from a technical university. This iterative refinement led to the establishment of design criteria for the respective conditions described in the following subsections.

### Boredom

Boredom is defined as "an unpleasant, transient affective state in which the individual feels a pervasive lack of interest in and difficulty concentrating on the current activity" (Fisher, 1993, p. 396). However, boredom in a game context can be seen as the counterpart to player engagement (which is supposedly elicited by the immersion and flow designs, described below). Seeing boredom as a relative experience at the lower end of a scale of engagement, we propose the following level design criteria for a *less-engaging* experience in a FPS game:

- Linear level layout (proceed through level on a line from start to end)
- Weak and similar opponents (e.g., only of two different enemy types)
- Repeating textures and models
- Damped and dull sounds

- No real winning or ending condition (after reaching the end of a level, the church, the player can continue to walk around)
- Limited choice of weapons and ammunition
- High amount of health and ammo supplies (for one weapon type) throughout the level
- No surprises (no gameplay information should be concealed)

## Immersion

According to our discussion in the introduction, we use immersion here as a description for the audiovisual or sensory experience of the game environment, which suggests the following immersion design criteria for a FPS game:

- Complex and exploratory environment with concealed information (player has to explore the area to find the way through the level)
- Various opponents (less and weak in the beginning, strong and numerous toward the end)
- Fitting sensory effects (fires, lighting, scripted animations, sounds, etc.)
- Variety of models, textures and dynamic lights to establish a mood and scenery
- New weapons are usually found after a fight as a reward as is ammo and health
- Narrative framing (we believe this would add to immersion, in our design it was however left out due to time limitations)

## Flow

The design criteria for flow are more concentrated on the sequence, pace and difficulty of challenges than on environmental settings. The FPS game level design guidelines that we used for the flow implementation are:

- Concentrate on the mechanics of one specific weapon and design the challenges around that (In our case, we ended up using the crossbow, which has a slow reload time, which is an interesting combat game mechanic.)
- Start with easy combat (Weak enemies are put in the start area with a slow spawn time, resulting in persistent but less challenging combat)
- Increase combat difficulty gradually (Combat becomes more difficult throughout the level as the number of opponents, attack pace and strength increases, while the spawn time decreases)
- Allow for half-cover spots. Between the areas of combat, we put short half-cover or rest spots, where players can find a sparse amount of health and ammo items. While these spots might save players from immediate attacks they do not provide a perfectly safe cover, so that the player must engage in combat again

In reality, not all of these criteria were equally well implemented, but here they serve as general guidelines in the FPS level design process. Each experiment participant played under each design condition in the same order. Physiological responses were measured as indicators of valence and arousal (Lang, 1995; Ravaja, 2004) together with questionnaires assessing self-reported game experience (IJsselsteijn, et al., 2008) and spatial presence (Vorderer, et al., 2004). Thus, physiological measurement of EDA and EMG were the dependent variables in this experiment together with subjective questionnaire responses.

### Participants

Data were recorded from 25 male University students, aged between 19 and 38 (*M* = 23.48, *SD* = 4.76). As part of the experimental setup, demographic data were collected with special respect to the suggestions made by Appelman (2007). Of the participants *88%* were right-handed. All the participants owned a personal computer (PC) and *96%* rated this as their favorite gaming platform. Other preferred platforms were Xbox 360, Playstation 3 and PS2. All participants played games at least twice a week, while *60%* play every day, *84%* played between two and four hours per day. The preferred mode of play was console single player (*44%*) or PC multiplayer (*36%*), while eight percent rated PC single player as their preferred play mode. 36% rated First-Person Shooters (FPS) as their favorite game type. Of the participants *44%* started to play digital games when they were younger than six years and *40%* started between six and eight years old. This leaves only *16%* that started to play between eight and twelve years. So, all the participants started playing digital games before twelve years. None of the subjects received any compensation for their participation in the experiment.

### Procedure

We conducted all experiments on weekdays with the first time slot beginning at 10:00 and the last ending at 20:00. General time for one experimental session was 2 hours with setup and cleanup. The experiments were advertized especially to graduate and undergraduate students. All participants were invited to a game research laboratory. After a brief description of the experimental procedure, each participant filled out two forms. The first one was a compulsory informed-consent form (with a request not to take part in the experiment when suffering from epileptic seizures or game addiction). The second form was an optional photographic-release form, which most of the participants signed as well. The participants were led to a notebook computer, where they filled out the initial game demographic questionnaire.

Participants were then seated in a comfortable office chair, which was adjusted according to their individual height. The electrodes were attached and participants were asked to relax. During this resting period of approximately 5 minutes, physiological baseline recordings were taken. Then, the participants played the game levels described above. Each game session was set to 10 minutes, but in general participants could finish all game levels before this. After each level, participants filled out a paper version of the game experience questionnaire (GEQ) to rate their experience. After completion of the experiment, all electrodes were removed. The participants were debriefed and escorted out of the lab.

### Measures

The following measurements were used:

> **Facial EMG**. We recorded the muscle activity from left *Orbicularis Oculi* (OO), *Corrugator Supercilii* (CS), and *Zygomaticus Major* (ZM) muscle regions (Fridlund & Cacioppo, 1986), using BioSemi flat-type active electrodes (11mm width, 17mm length, 4.5mm height) with sintered Ag-AgCl (silver/silver chloride) electrode pellets having a contact area 4 mm in diameter. The electrodes were filled with low-impedance, highly conductive Signa electrode gel (Parker Laboratories, Inc.). The raw EMG signal was recorded with the ActiveTwo AD-box at a sampling rate of 2 kHz and using ActiView acquisition software.

**Electrodermal activity**. Electrodermal activity (i.e., impedance of the skin) was measured using two passive Ag-AgCl (silver/silver chloride) Nihon Kohden electrodes (1μA, 512 Hz). The electrode pellets were filled with Signa electrode gel (Parker Laboratories, Inc.) and attached to the thenar and hypothenar eminences of a participant's left hand (Boucsein, 1992).

**Video recording**. A Sony DCR-SR72E PAL video camera (handycam) was put on a tripod and positioned approximately 50 cm behind and slightly over the right shoulder of the player for observation of player movement and in-game activity. In addition, the video recordings served as a validation tool when psychophysiological data were visually inspected for artifacts and recording errors.

**Gameplay experience and spatial presence questionnaire**. Different components of game experience were measured using the game experience questionnaire (GEQ) (IJsselsteijn, et al., 2008). The questionnaire was developed on the basis of focus group research (Poels, de Kort, & IJsselsteijn, 2007) and following investigations among frequent players. It consists of the seven dimensions *flow*, *challenge*, *competence*, *tension*, *negative affect*, *positive affect* and *sensory and imaginative immersion* that are measured each using 5-6 questionnaire items in the long version. Each item consists of a statement on a five-point scale ranging from 0 (not agreeing with the statement) to 4 (completely agreeing with the statement). As shown in a previous assessment by Nacke and Lindley (2008), the GEQ components can assess experiential constructs with good reliability.

In addition, we employed the MEC Spatial Presence Questionnaire (SPQ) (Vorderer, et al., 2004). More precisely, we used the spatial presence: self location (SPSL) and spatial presence possible actions (SPPA) subscales, each measured with four items. Each item consisted of a statement on a five-point scale ranging from 1 ("I do not agree at all") to 5 ("I fully agree"). Other apparatus used but not included in this analysis were a Biosemi 32-channel EEG system and a Tobii 1750 eye tracker. This additional data will form the basis of future papers.

### Data Reduction and Analysis

Recorded psychophysiological data were inspected visually using BESA (MEGIS Software GmbH, Germany) software to check correctly recorded signals. To reduce noise, EMG data were also filtered using a low cutoff filter (30 Hz, Type: forward, Slope: 6 dB/oct) and a high cutoff filter (400 Hz, Type: zero phase, Slope: 48dB/oct). If data remained indistinct, they were excluded from further analysis. Tonic EMG data were rectified and exported together with tonic EDA data at a sampling interval of 0.49 ms to SPSS software (SPSS Inc.) for further statistical analysis. EMG data was transformed with a natural logarithm to reduce skew. Descriptive statistics were calculated for each person over the complete game session. EDA data was log-transformed for normalization.

**Results**

*GEQ Results*

For assessing dimensions of game experience, the GEQ was used (IJsselsteijn, et al., 2008). The comparison of average scores is shown in Figure 3. Average scores and reliability of these results have been briefly discussed by Nacke and Lindley (2008). The notable results are an increase in positive affect and immersion for the immersion level. Accordingly, this level scores lowest for negative affect items. The boredom level scores lowest on challenge, immersion and flow, but highest on competence, which is completely in line with expectations. The flow level scores lowest on competence, but highest on flow, challenge and tension, which the following analysis will support to be the most significant result.

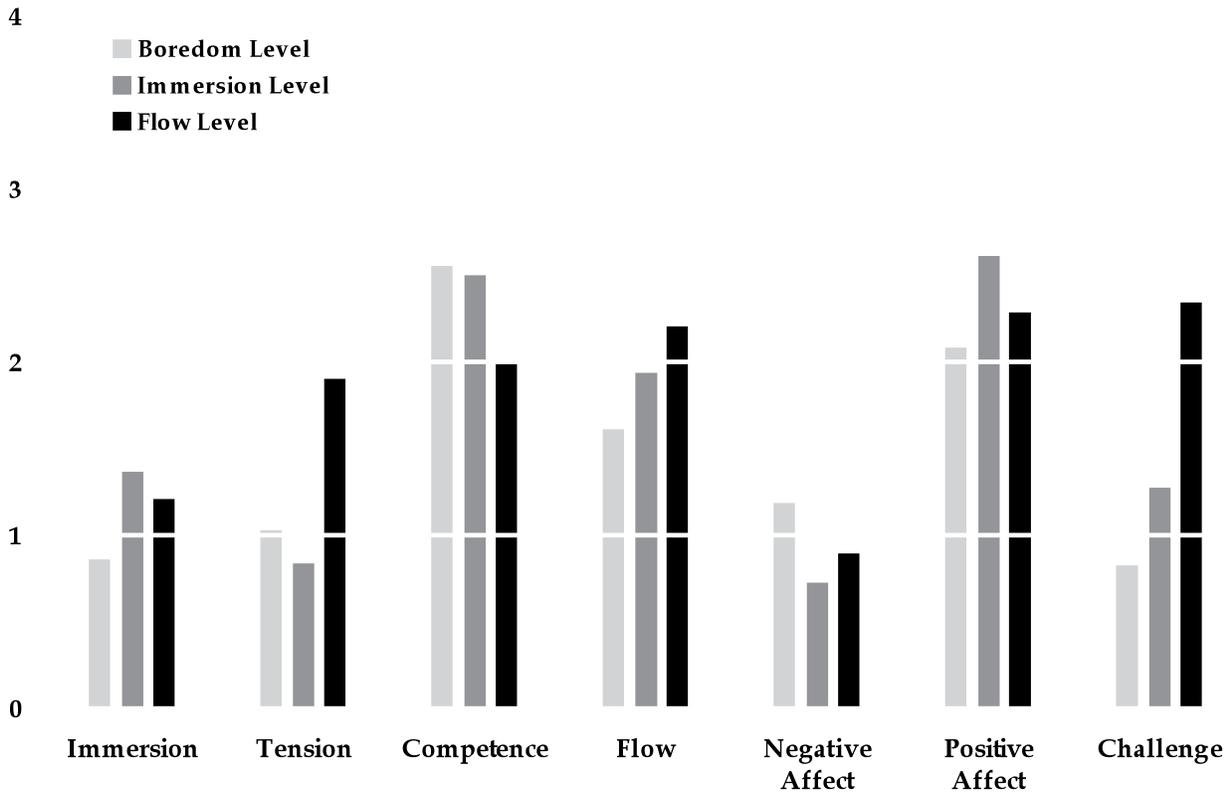

*Figure 3: Mean scores for GEQ components in each level (scale from 0 to 4). For more validity statistics, please see Nacke and Lindley (2008).*

To test statistical significance of the results, one-way repeated-measures analyses of variance (ANOVAs) were conducted in SPSS using the game mod levels as the within-subject factor for each measurement. For GEQ components immersion ($\chi^2(2) = 1.13$, $p > .05$), flow ($\chi^2(2) = 1.15$, $p > .05$), positive affect ($\chi^2(2) = 0.16$, $p > .05$), negative affect ($\chi^2(2) = 0.66$, $p > .05$), challenge ($\chi^2(2) = 2.96$, $p > .05$) and tension ($\chi^2(2) = 4.68$, $p > .05$), Mauchly's test indicated that the assumption of sphericity had been met. For the remaining component competence ($\chi^2(2) = 10.72$, $p < .05$) it was violated. Therefore, degrees of freedom were corrected for the competence component using Greenhouse-Geisser estimates of sphericity ($\varepsilon = .70$).

Statistical significance was unfortunately not achieved for the components: Immersion: $F(2, 40)$ = 2.00, $p > .05$), competence: $F(1.40, 27.95) = 2.34$, $p > .05$), flow: $F(2, 40) = 2.08$, $p > .05$), positive affect: $F(2, 40) = 1.94$, $p > .05$), and negative affect: $F(2, 40) = 1.90$, $p > .05$). The items challenge: $F(2, 40) = 32.54$, $p < .05$) and tension: $F(2, 40) = 7.98$, $p < .05$) were both clearly statistically significant. This is a sign of the subjective game experiences "challenge" and "tension" (measured with the GEQ) being significantly affected by the different gameplay experience modalities.

### Spatial Presence Results

Table 1 shows the mean scores for the MEC Spatial Presence Questionnaire (Vorderer, et al., 2004). It can be noted that spatial presence possible actions ratings were highly increased in the level designed for immersion. Spatial presence scores are lowest in the boredom level.

| Design Focus | Spatial Presence Self-Location | Spatial Presence Possible Actions |
|---|---|---|
| Boredom | 2.07 (1.10) | 2.57 (1.06) |
| Immersion | 2.60 (0.96) | 3.30 (0.85) |
| Flow | 2.68 (1.22) | 2.62 (1.12) |

*Table 1: Means (and standard deviations) of the MEC Spatial Presence Questionnaire (scale from 1 to 5) by Vorderer, et al. (2004).*

For Spatial Presence components Self-Location ($\chi^2(2) = 4.73$, $p > .05$) and Possible Actions ($\chi^2(2) = 2.73$, $p > .05$) Mauchly's test indicated that the assumption of sphericity had been met. In addition, statistical significance was achieved for both components, Possible Actions: $F(2, 40)$ = 4.79, $p < .05$) and Self-Location: $F(2, 40) = 3.40$, $p < .05$). These results show that the subjective feeling of spatial presence was significantly affected by the different gameplay experience modalities.

### EMG Results

Electromyographic (EMG) Activity per Level (ln[μV])

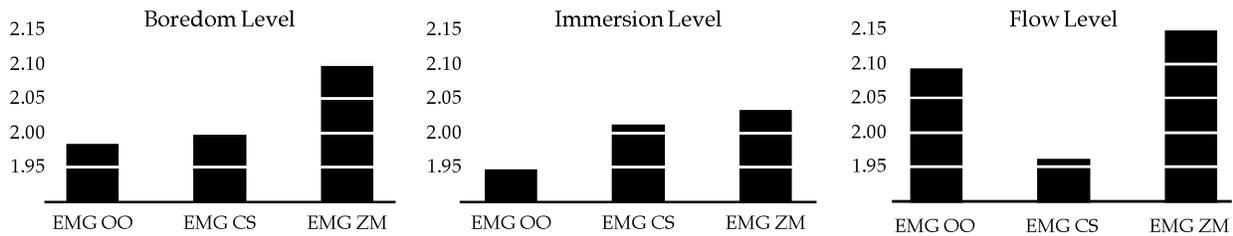

*Figure 4: EMG activity for each level design focus. EMG activity is displayed as ln[μV].*

Figure 4 displays the cumulative averages over the playing time for all participants in all levels. Histograms for EMG measures were visually inspected and data was assumed to be normally distributed. EDA data was logarithmically normalized. Positively valenced emotions would be indexed by increased ZM and OO activity (Ravaja, et al., 2008). The game level designed for the flow condition shows the highest values for positive valence (measured by OO and ZM activity) as well as for arousal (i.e., EDA). In contrast to this, the immersion level scores lowest on valence as well as arousal. The boredom level scores similar, but a bit above the values for the immersion level for all physiological measurements except CS activity (i.e., negative valence). Mauchly's test indicated that the assumption of sphericity had been met for EMG activity in OO ($\chi^2(2) = 0.60$, $p > .05$), CS ($\chi^2(2) = 3.33$, $p > .05$) and ZM ($\chi^2(2) = 4.32$, $p > .05$) muscle regions.

A one-way repeated-measures ANOVA was conducted using level design (boredom, immersion, flow) as a three-level within-subject factor for dependent variables OO, CS and ZM EMG activity (ln[μV]). Multivariate tests showed a significant impact of level design on EMG activity, $F(6, 10) = 8.08$, $p < .01$. However, CS EMG activity showed no significant difference in the level designs, OO EMG activity was only marginally significant, $F(2, 30) = 3.09$, $p = .06$, but ZM EMG activity was significantly affected by the level design, $F(2, 30) = 6.65$, $p < .01$. Polynomial contrasts showed a significant quadratic trend for OO EMG activity, $F(1, 15) = 6.11$, $p < .05$, and ZM EMG activity, $F(1, 15) = 11.12$, $p < .01$. A follow-up test with repeated contrasts revealed a significant difference in OO EMG activity between boredom level design and flow level design, $F(1, 15) = 6.05$, $p < .05$, as well as significant differences in ZM EMG activity between boredom level design and flow level design, $F(1, 15) = 8.90$, $p < .01$, and flow level design and immersion level design, $F(1, 15) = 9.27$, $p < .01$. Interestingly, ZM elicits most activity compared to the other muscles, but is lowest in the immersive level and highest in the flow level. The EMG results show that objective physiological responses (for all measures taken except in the CS region) from an accumulated game session were significantly influenced by the different gameplay experience modalities.

### EDA Results

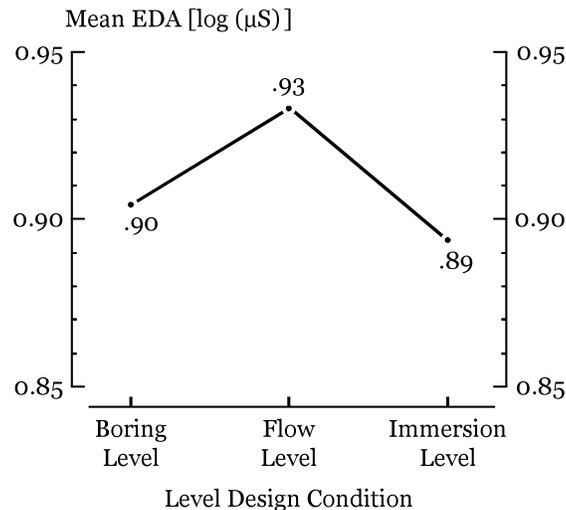

*Figure 5: EDA in the level designs. EDA is displayed as log[μS]*

A one-way repeated-measures ANOVA was conducted on the log-transformed electrodermal activity data (log[μS]). For these data, sphericity was violated ($\chi^2(2) = 10.14$, $p < .05$) and corrected using Greenhouse-Geisser estimates ($\varepsilon = .66$). Following this, a significant impact of level design on EDA could be found, $F(1.32, 19.80) = 4.34$, $p < .05$. Polynomial tests of within-subjects contrasts showed a significant quadratic trend, $F(1, 15) = 6.94$, $p < .05$, which was followed up with repeated contrasts revealing a significant difference between boredom level and flow level, $F(1, 15) = 12.09$, $p < .01$. While EDA was almost equal for boredom level and immersion level (see Figure 5), it was significantly increased during the flow level. Thus, the flow level was physically more arousing to play than the other levels.

## Discussion and Future Work

This paper has described and analyzed the results of an experiment to measure gameplay experience and its effect on player valence and arousal. It was the goal as well to detect any possible correlations between measurable valence and arousal features and self-reported subjective experience.

To begin with, the GEQ showed that it could accurately measure its components, but that only challenge and tension showed significant discrimination in this experiment. This can be due to the fact that in the design of the game levels, we relied on subjective experience and iterative feedback to design for each concept: boredom, immersion and flow. While flow and boredom might be intuitively understood by most gamers, immersion certainly is not. The challenge aspect of flow seems to be the one best assessed with the GEQ as it shows a high increase in the flow level (which had gradually increasing combat challenges throughout the level). This of course leads to this level culminating in a very challenging end fight and thus might have been perceived as holistically more challenging, even though combat at the start of the level had the same density as in the immersion level. Overall, the GEQ results seem to validate the intended level design for the flow level. However, there seems not to be enough evidence in the data to subjectively discriminate between experiences in the immersion and the boredom levels. This might be an indicator that flow is a much better understood concept for level design than for example immersion or boredom. From personal experience, we can say that it was much easier to come up with design guidelines for the flow level than with those for the immersion level.
The measurements of spatial presence appear to be more interesting. The level designed for immersion scores high on "self-location" and highest on "possible actions". Thus, it is very likely that we subjectively designed for what Ermi and Mäyrä (2005) would call *imaginative immersion* and that this feeling might be related to *spatial presence*, especially in the dependency of presence upon what Vorderer et al. (2004) describe as "possible actions". In contrast to this, the feeling of "self-location" might be directly linked to flow in combat experiences, since the flow level scores higher than the immersion level in this item. Clearly, these results present once again the need to find a more distinct terminology for the different forms of immersion and presence.

Finally, the measurement of EMG responses was significant for the muscles indicating positive valence (OO and ZM). In addition, the measurement of arousal (EDA) showed statistically significant differences under the different conditions manifest in the different level designs. The flow level scores highest under these conditions, making it a foundation for high-arousal, positive emotions. We consider this a noteworthy finding, because it supports the link between gradual challenges in a competitive environment and positive emotions. Our initial assumption before we created the design guidelines was slightly biased to the contrary, assuming that highly challenging gameplay could be frustrating and could leave players with a negative feeling. However, according to our results and in line with our design hypothesis (formed based on the level design guidelines), the opposite is true: Challenging levels are experienced as being more arousing and deliver more positive emotions than boring levels. Joy in this case does not come from victory or success, but from challenging gameplay (Ravaja, et al., 2005; Ravaja, Saari, Salminen, Laarni, & Kallinen, 2006; Ravaja, et al., 2008).

The psychophysiological findings contradict the findings of Kivikangas (2006) that EMG activity over ZM and OO (positive valence) does not support a relationship with flow. If we assume that we can accurately assess flow with the GEQ (IJsselsteijn, et al., 2008; Nacke, Nacke, & Lindley, 2009), then it is supported in our study to be related to positive emotion as indexed by physiological responses. A caveat of this study is that it was focused on male hardcore gamers only and thus it might be hypothesized that these results are only valid for this target group. It remains for future research to indicate whether psychophysiological measurements can accurately describe gameplay experiences for a broader demographic population. At this point, we have also not conducted a statistical correlation of constructs with physiological measures, which will also certainly give more useful results in the future.

In considering the limitations of the experiment design described here, it may be proposed that future research might explore different time resolutions[3], since emotional responses to a complete play session might be linked to smaller scale details of the modulation of emotional reactions over a sequence of specific game events (Nacke, Lindley, & Stellmach, 2008; Ravaja, et al., 2008; Salminen & Ravaja, 2008), such as player death events.

In conclusion, the study reported here supports that physiological responses can be an indicator of psychological player states in gameplay experience as indicated by relation to subjective player reports. Ongoing work will investigate the relationship between physiological responses and subjective experiences in greater detail.

## Acknowledgements

The research reported in this paper has been supported by the EU FP6 FUGA (Fun of Gaming) research project (NEST-PATH-028765). We thank our FUGA colleagues, especially Niklas Ravaja, Matias Kivikangas, Simo Järvelä, Wouter van den Hoogen, and Karolien Poels as well as Anders Drachen and Mark Grimshaw for great support and stimulating discussions. We would also like to thank Dennis Sasse and Sophie Stellmach for their help with the level designs.

Zahorik, P., & Jenison, R. L. (1998). Presence as Being-in-the-World. *Presence: Teleoperators and Virtual Environments, 7*(1), 78-89. doi: 10.1162/105474698565541

---

[1] Both in correlating the degree of emotions experienced and comparing the session scale of subjective reports with the millisecond scale of psychophysiological recordings of emotional changes and modulations

[2] It must be noted that the experiment has a preliminary nature due to the fact that the psychophysiological characterization of gameplay experiences (such as immersion or flow) is not well developed, yet.

[3] Rather than looking at the net effect of subjectively reported experience over a complete play session.